\documentclass{article}

\usepackage{amsmath,amssymb}

\usepackage{arxiv}
\usepackage{orcidlink}
\usepackage[utf8]{inputenc} 
\usepackage[T1]{fontenc}    
\usepackage{hyperref}       
\usepackage{url}            
\usepackage{booktabs}       
\usepackage{amsfonts}       
\usepackage{nicefrac}       
\usepackage{microtype}      
\usepackage{lipsum}
\usepackage{graphicx}
\graphicspath{ {./images/} }

\usepackage{listings} 
\usepackage{xurl} 

\title{A note on closed-form solutions for estimating sample size when externally validating a binary prediction model based on $C$-statistic precision}

\author{
 Denis A. Shah \\
 Department of Plant Pathology\\
 Kansas State University\\ 
 Manhattan, KS 66506\\
  \texttt{dashah81@ksu.edu} \\
  \And
 Erick D. De Wolf \\
  Department of Plant Pathology\\
 Kansas State University\\ 
 Manhattan, KS 66506\\
  \texttt{dewolf1@ksu.edu} \\
  \And
 Pierce A. Paul \\
  Department of Plant Pathology\\
  The Ohio State University CFAES\\
  Wooster, OH 44691 \\
  \texttt{paul.661@osu.edu} \\
  \AND
  Laurence V. Madden \\
  Department of Plant Pathology\\
  The Ohio State University CFAES\\
  Wooster, OH 44691 \\
  \texttt{madden.1@osu.edu}
}

\begin{document}
\maketitle
\begin{abstract}
External validation of clinical prediction models is crucial for assessing whether they are fit for use. The $C$-statistic is a widely used measure of discriminative performance of such models predicting a binary outcome. A method for obtaining the minimum sample size required for the precise estimation of the $C$-statistic during validation, based on the rearrangement of Newcombe's formula for the standard error of the $C$-statistic \{SE($C$)\}, was recently proposed and implemented in R and Stata software via an iterative computational approach. We present seven novel closed-form solutions, derived using different computer algebra systems and artificial intelligence models, to the algebraic rearrangement of Newcombe's formula. We present these distinct forms to demonstrate how different computational tools yield structurally distinct but mathematically equivalent solutions, and to evaluate their practical differences in computational performance. Our closed-form solutions yield identical sample size estimates to the iterative method when applied to illustrative examples. In a benchmarking analysis, the closed-form solutions were on average 148,000 to 264,000 times faster in median execution time than the current iterative implementation, while also exhibiting minor efficiency differences among themselves. This work provides a validated, highly efficient computational tool applicable to sample size calculation for external validation studies. R code functions implementing the closed-form solutions are provided.
\end{abstract}


\section{Introduction}
Prediction models for a binary outcome are used heavily in medical practice as diagnostic and prognostic aids and are an integral part of clinical decision-making~\cite{collins_tripodai_2024,efthimiou_developing_2024,shipe_developing_2019}. If these models are to be considered fit for purpose, their performance must be objectively evaluated on independent datasets not used during the model development process; in other words, they should undergo some form of external validation~\cite{collins_evaluation_2024,tiruneh_externally_2024}. Proper external validation requires a sufficient sample size to precisely estimate measures of model performance~\cite{riley_minimum_2021,riley_evaluation_2024}. 

Riley et al.~\cite{riley_minimum_2021} presented formulas for determining the minimum external validation sample size required to precisely estimate calibration, discrimination and clinical utility of a binary prediction model. For the discrimination criterion, they proposed using Newcombe's formula~\cite{newcombe_confidence_2006} for the standard error of the $C$-statistic. For binary outcomes, the $C$-statistic is equivalent to the Area Under the Receiver Operating Characteristic Curve (AUROC) and represents the probability that the model assigns a higher risk to a case than a non-case. It is a standard metric for quantifying the discriminative ability of clinical prediction models. However, they relied on an iterative numerical solution to calculate the validation sample size, since a direct algebraic rearrangement of Newcombe's formula~\cite{newcombe_confidence_2006} was deemed “not possible” in their original contribution \cite{riley_minimum_2021}, which was also restated in a later paper \cite{riley_evaluation_2024}. The iterative approach is implemented in their accompanying R and Stata code (\url{https://github.com/JoieEnsor}). 

 In this note, we demonstrate that closed-form algebraic rearrangements of Newcombe's formula~\cite{newcombe_confidence_2006} to calculate the validation sample size are indeed possible. We present derivations obtained manually with the aid of computer algebra systems, which assist in streamlining complex calculations, and with prompts passed to different artificial intelligence (AI) products to take advantage of recent advances in their algebraic problem-solving capabilities. We then empirically verify that our closed-form solutions yield sample size estimates identical to those obtained via the iterative approach, illustrated with examples provided by Riley et al.~\cite{riley_minimum_2021}. Finally, we conduct an evaluation timing benchmark that highlights the substantial efficiency gains offered by the closed-form solutions. The current paper expands on a brief rapid response note pointing to the fact that a closed-form solution was possible \cite{shah_rapid_2024}. 

 \section{Background}
 Riley et al.~\cite{riley_minimum_2021} discussed the reasons for preferring the formula for the standard error of the $C$-statistic proposed by Newcombe~\cite{newcombe_confidence_2006} over other formulations, when considering the external validation sample size needed for the precise estimation of the $C$-statistic. Specifically, Newcombe’s formula is derived from nonparametric principles, making it independent of distributional assumptions on the underlying model’s linear predictor function and it performs well based on empirical evaluations.

Newcombe’s formula was presented by Riley et al. as equation 11 in their paper~\cite{riley_minimum_2021}: 
\begin{equation}
\label{eq:eq1}
{\text{SE}(C)} = \sqrt{\frac{C(1-C)\left(1+ \left(\frac{N}{2}-1\right) \left(\frac{1-C}{2-C}\right)+\frac{\left(\frac{N}{2}-1\right)C}{1+C}\right)}{N^2\phi  (1-\phi)}}
\end{equation}
where SE($C$) is the standard error of the $C$-statistic ($C$), $N$ is the number of individuals (sample size) from the target population (i.e., the validation set) and $\phi$ is the anticipated outcome event proportion in the validation population.

\section{Derivation of closed-form solutions for \texorpdfstring{$\boldsymbol{N}$}{N}}
Eq~(\ref{eq:eq1}), is daunting to rearrange by hand to solve for $N$ as a function of $C$ (the $C$-statistic), $\phi$ (the outcome event proportion) and SE($C$) (the standard error of the $C$-statistic). To explore whether the problem is solvable, we utilized two distinct categories of computational tools: Computer Algebra Systems (CAS), which are deterministic software programs designed for symbolic manipulation, and Large Language Models (LLMs), which are non-deterministic, generative AI systems. We used both a commercial (Mathematica version 12.0) and an open-source (Maxima version 23.05.1) CAS product. 

A note on notation: to maintain fidelity to the outputs generated by the various computational tools and minimize transcription errors, we have retained the specific intermediate variable names (e.g., $\alpha, \beta, \gamma$) used by each system. Consequently, these symbols are \textit{locally defined} within each solution and do not represent consistent quantities across the different equations.

\subsection{With Mathematica}
Eq~\ref{eq:eq1} was returned in a simpler form after squaring both sides and getting rid of the fractions in the numerator:
\begin{equation}
\label{eq:eq2}
{\text{SE}(C)}^2 = \frac{(C-1)C\, \left\{2(C-1)C(N-1) - (N+2)\right\}}
     {2(C-2)(C+1)N^2\, (\phi-1)\phi}
\end{equation}

Solving Eq~\ref{eq:eq2} for $N$ gave two solutions: 
\begin{subequations}
\label{eq:eq3}
\begin{align}
N = \frac{C+C^2-4C^3+2C^4-Z}{4 {\text{SE}(C)}^2 (C-2) (C+1) (\phi-1) \phi} \label{eq:eq3a} \\
N = \frac{C+C^2-4C^3+2C^4+Z}{4 {\text{SE}(C)}^2 (C-2) (C+1) (\phi-1) \phi} \label{eq:eq3b}
\end{align}
\end{subequations}
where,
\begin{equation*}
Z = \Biggl[ 
   \bigl( C - 1 \bigr) C 
   \Biggl( 
      \bigl( C - 1 \bigr) C \, 
      \bigl( 1 - 2 (C - 1) C \bigr)^{2} 
      + 16 \, {\text{SE}(C)}^2 
      \bigl( -2 + C - 2C^{3} + C^{4} \bigr) 
      \phi \, (1-\phi) 
   \Biggr) 
\Biggr]^{1/2}
\end{equation*}

$N$ is of course strictly positive. That the correct solution was given by Equation~\ref{eq:eq3b} was confirmed by substituting values for $C$, $\phi$ and ${\text{SE}(C)}^2$; for example, $C = 0.7$, $\phi= 0.1$ and ${\text{SE}(C)}^2={0.02551}^2$ gave $N = -1.1116$ by Eq~\ref{eq:eq3a} and $N = 1153.03$ by Eq~\ref{eq:eq3b}. 

Visual inspection of the structure of Eq~\ref{eq:eq3b} revealed repeated subexpressions. Making the further substitutions $\alpha=C-1$, $\beta=16\left(-2+C-2C^3+C^4\right){\text{SE}(C)}^2\phi$ and $\mu=C\left(C-1\right)=C\alpha$ we arrived at
\begin{equation}
\label{eq:eq4}
N= \frac{C + C^2 - 4C^3 + 2C^4 + \sqrt{\mu \left(\beta+(1-2\mu)^2\mu -\beta\phi \right)}}{4{\text{SE}(C)}^2 (\alpha-1)(\alpha+2)(\phi-1) \phi}
\end{equation}
One final substitution, $\delta=C+C^2-4C^3+2C^4$, led to a compact closed-form formula for $N$: 
\begin{equation}
\label{eq:eq5}
N = \frac{\delta + \sqrt{\mu \left(\beta + (1-2\mu)^2 \mu - \beta\phi\right)}}{4{\text{SE}(C)}^2 (\alpha-1)(\alpha+2)(\phi-1)\phi}.
\end{equation}

\subsection{With Maxima}
We followed the same procedural steps as used with Mathematica (i.e., squaring the expression, simplifying the representation, root finding, testing for positivity, pattern recognition, and substitution). The first two steps gave the following expression for ${\text{SE}(C)}^2$:
\begin{equation}
\label{eq:eq6}
{\text{SE}(C)}^2 = \frac{(C-1)C\left(2C^2N-2CN-N-2C^2+2C-2\right)}{2(C-2)(C+1)N^2(\phi-1)\phi}
\end{equation}
which is the same as Eq~\ref{eq:eq2} but with the numerator term after $\left(C-1\right)C$ fully expanded. 

Solving Eq~\ref{eq:eq6} for $N$ gave two roots (not shown because of their length) as before, with the correct (positive) solution identified after substituting values for $C$, $\phi$ and ${\text{SE}(C)}^2$. Repeated patterns were again evident, and after making the substitutions $w=16C^6-48C^5+32C^4+16C^3-48C^2+32C=16C(C-1)(C-2)(C+1)(C^2-C+1)$, $x=\phi{\text{SE}(C)}^2$ and $y=2C^4-4C^3+C^2+C=C(C-1)(2C^2-2C-1)$ we arrived at a compact solution for $N$:
\begin{equation}
\label{eq:eq7}
N = \frac{y+\sqrt{-(wx\phi)+y^2+wx}}{4 (C-2) (C+1) x (\phi-1)}
\end{equation}
It is straightforward to prove that the denominators of Eqs~\ref{eq:eq5} and \ref{eq:eq7} are the same algebraically. 

\section{Artificial intelligence solutions}
We used the following LLMs in finding closed-form solutions for $N$: Perplexity’s Sonar, Anthropic’s Claude Sonnet 4.0, OpenAI’s GPT 4.1, and xAI’s Grok 3 Beta. Each of these models was specified as options within Perplexity Pro. We also used Claude Sonnet 4.0, Google’s Gemini 2.5 Pro and MathGPT (\url{https://math-gpt.org/}) directly. Each AI model was given the following prompt, in which equation \ref{eq:eq2} was presented in \LaTeX format: 
\begin{lstlisting}[breaklines=true]
Generate a closed-form solution for N from the equation:
{SE}^2 = \frac{C(1-C)\left(1+ \left(\frac{N}{2}-1\right) \left(\frac{1-C}{2-C}\right)+\frac{\left(\frac{N}{2}-1\right)C}{1+C}\right)}{N^2\phi (1-\phi )}. Assume N>0. Identify and substitute repeated patterns with new variables (e.g., \alpha, \beta) to achieve a compact representation. Present the final solution for N in LaTeX. Confirm the positivity of the solution by substituting the values {SE}^2 = 0.02551^2, C = 0.7, and \phi = 0.1.
\end{lstlisting}
Prompts were run on July 10\textsuperscript{th} and 11\textsuperscript{th} 2025; we documented the dates because of the rapidly evolving nature of AI models. 

Most AIs tried finding an intermediate expression that would lead to a quadratic equation, which would permit using the quadratic formula to solve for $N$. All AIs generated solutions for $N$; not all were correct as confirmed by numeric substitution of the values of ${\text{SE}(C)}^2$, $C$ and $\phi$ specified in the prompt. The correct value for $N$ given ${\text{SE}(C)}^2= 0.02551^2$, $C = 0.7$ and $\phi=0.1$ is 1153.03. 

We encountered nuances which only served to illustrate that close checking by hand plus confirmation by numeric substitution of values for ${\text{SE}(C)}^2$, $C$ and $\phi$ was required with any AI solution. For example, repeating the prompt using the same AI model sometimes gave different answers. GPT 4.1 and Grok 3 Beta gave algebraically correct solutions but got the math wrong in numerically confirming the solutions. Claude Sonnet 4.0 used directly gave a solution with algebraic steps that \textit{seemed} plausible without double-checking the accuracy by hand. However, it was confirmed incorrect by Claude Sonnet 4.0’s own numeric substitutions and by our coding of the provided solution. On the other hand, Claude Sonnet 4.0 called within Perplexity gave a complicated solution with unclear and illogical steps, which we did not pursue further. These results were not surprising as AIs do struggle with more complex algebraic problems~\cite{spreitzer_mathematical_2024,valverde_wolframalpha_2025}. Correct (numerically verified) closed-form solutions for $N$ were given by Sonar, GPT 4.1, Grok 3 Beta, Gemini 2.5 Pro and MathGPT.

\subsection{Sonar's solution}
Sonar returned the solution
\begin{equation}
\label{eq:eq8}
\begin{split}
N = \frac{1}{D} \Biggl( & C(1-C)(\alpha + \beta) \\
& + \sqrt{\left\{C(1-C)(\alpha + \beta)\right\}^2 + 4D \cdot C(1-C)\left\{1 - (\alpha + \beta)\right\}} \Biggr)
\end{split}
\end{equation}
where $D = 4 {\text{SE}(C)}^2 \phi (1-\phi)$, $\alpha=\frac{1-C}{2-C}$ and $\beta=\frac{C}{1+C}$. 

\subsection{GPT 4.1's solution}
GPT 4.1 came up with
\begin{equation}
\label{eq:eq9}
N = \frac{\frac{C(1-C)\gamma}{2} + \sqrt{\left(\frac{C(1-C)\gamma}{2}\right)^2 + 4 {\text{SE}(C)}^2 \phi (1-\phi) C(1-C)(1-\gamma)}}{2 {\text{SE}(C)}^2 \phi (1-\phi)}
\end{equation}
where $\gamma=\frac{1-C}{2-C}+\frac{C}{1+C}$. It becomes immediately clear that GPT 4.1 combined the $\alpha$ and $\beta$ substitutions used by Sonar into $\gamma$, but failed to simplify the fractions in the numerator as Sonar did.

\subsection{Grok 3 Beta's solution}
Grok 3 Beta’s take on the solution was
\begin{equation}
\label{eq:eq10}
\begin{split}
N = \frac{1}{D} \Biggl( & \sqrt{C(1-C)} \sqrt{C(1-C) \gamma^2 - 4D(\gamma - 1)} \\
& + C(1-C) \gamma \Biggr)
\end{split}
\end{equation}
where, as with Eq~\ref{eq:eq8}, $D = 4 {\text{SE}(C)}^2 \phi (1-\phi)$ and, as with GPT 4.1's solution (Eq~\ref{eq:eq9}), $\gamma = \frac{1-C}{2-C} + \frac{C}{1+C}$. That is, Grok 3 Beta chose to factor out $\sqrt{C\left(1-C\right)}$ from the main square root term in the numerator compared with GPT 4.1’s solution. 

\subsection{Gemini 2.5 Pro’s solution}
Gemini 2.5 Pro returned the solution 
\begin{equation}
\label{eq:eq11}
N = \frac{\gamma \beta + \sqrt{\gamma^2 \beta^2 + 16\alpha\gamma(1-\beta)}}{4\alpha}
\end{equation}
where $\alpha={\text{SE}(C)}^2\phi\left(1-\phi\right)$, $\beta=\frac{1+2C-2C^2}{\left(2-C\right)\left(1+C\right)}$ and $\gamma=C\left(1-C\right)$. Eq~\ref{eq:eq11} is just a more compact representation of Eqs~\ref{eq:eq8} to \ref{eq:eq10}.

\subsection{An elegant solution with MathGPT}
The solution provided by MathGPT when prompted was logical, clear and elegantly simple. MathGPT first recognized the patterns in the squared version of Eq~\ref{eq:eq1} and made the substitutions $\alpha=\frac{N}{2}-1$, $\beta=\frac{1-C}{2-C}$, and $\gamma=\frac{C}{1+C}$, thereby simplifying the numerator bracketed term $\left(1+\left(\frac{N}{2}-1\right)\left(\frac{1-C}{2-C}\right)+\frac{\left(\frac{N}{2}-1\right)C}{1+C}\right)$ to $1+\alpha\beta+\alpha\gamma=1+\alpha\left(\beta+\gamma\right)$. The expression for ${\text{SE}(C)}^2$ then became
\begin{equation}
\label{eq:eq12}
{\text{SE}(C)}^2=\frac{C\left(1-C\right)\left\{1+\alpha\left(\beta+\gamma\right)\right\}}{N^2\phi\left(1-\phi\right)}
\end{equation}
Rearranging Eq~\ref{eq:eq12} for $N$ and making the further substitutions $\delta=\beta+\gamma$ and $A=\frac{C\left(1-C\right)}{{\text{SE}(C)}^2\phi\left(1-\phi\right)}$ led to a quadratic equation in $N$:
\begin{equation}
\label{eq:eq13}
N^2-\frac{A\delta}{2}N-A\left(1-\delta\right)=0
\end{equation}
which is of the general quadratic form $ax^2+bx+c=0$, where $a$, $b$ and $c$ are constants and $a\neq0$. The quadratic formula provides the solution to $x$ given the general quadratic form: $x=\frac{-b\pm\sqrt{b^2-4ac}}{2a}$. Applying the quadratic formula to Eq~\ref{eq:eq13} and checking the positivity requirement on $N$ led to the solution 
\begin{equation}
\label{eq:eq14}
N = \frac{A\delta + \sqrt{A^2\delta^2 + 16A(1-\delta)}}{4}
\end{equation}
where $A$ and $\delta$ were defined as above. 


\section*{Confirmation of closed-form solutions}
\begin{figure}[htbp]
\centering
\includegraphics[width=0.8\textwidth]{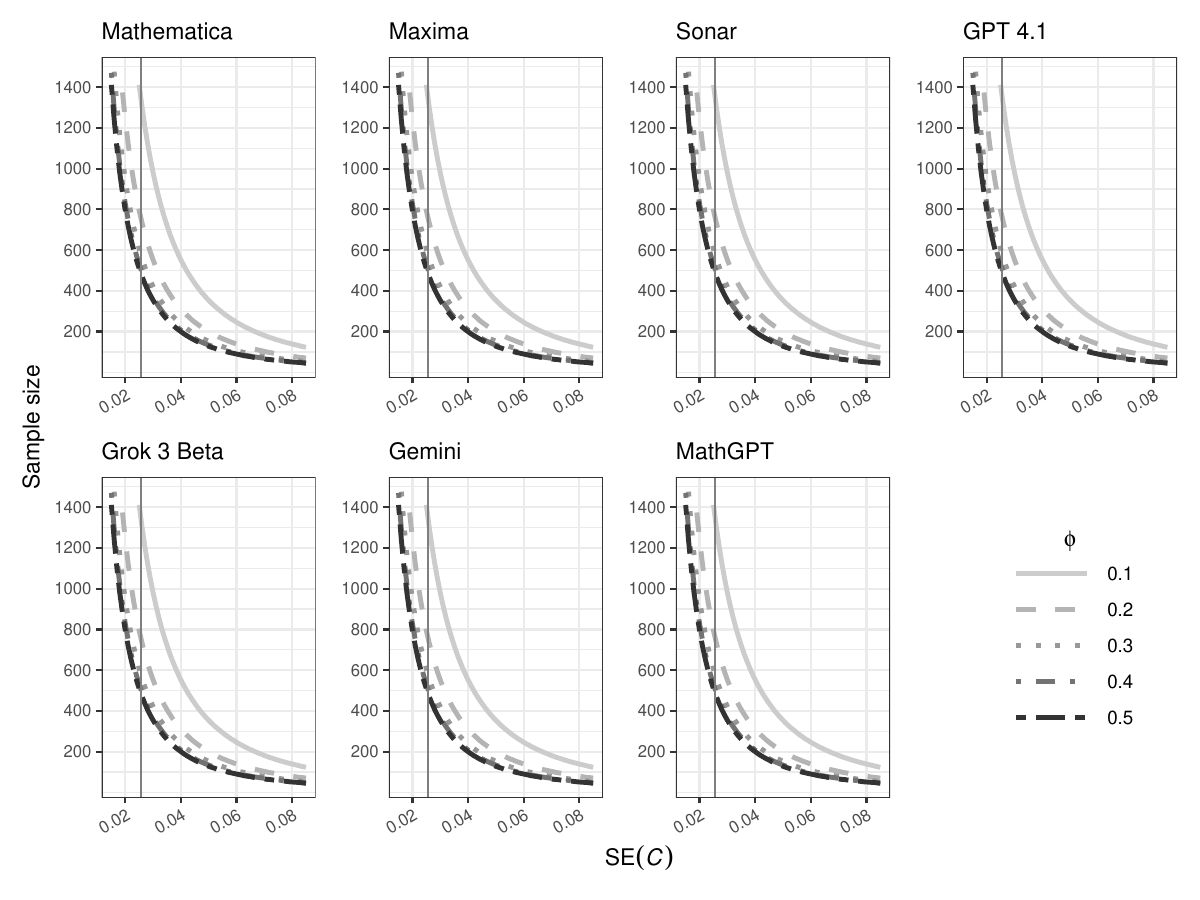}  
\caption{Values of the standard error of the $C$-statistic $\{{\text{SE}(C)\}}$ and estimated sample size calculated for $C = 0.6$ and outcome event proportions ($\phi$) of 0.1 to 0.5, for seven different closed-form solutions for the sample size required for the external validation of a binary prediction model.
The solid vertical line references ${\text{SE}(C)} = 0.02551$, corresponding to a confidence interval width of 0.1 for the $C$-statistic.}
\label{fig1}
\end{figure}

\begingroup
\sloppy 
In their paper, Riley et al.~\cite{riley_minimum_2021} gave five examples of the estimation of $N$ based on the ${\text{SE}(C)}$ criterion (Table~\ref{table-1}). They solved Eq~\ref{eq:eq1} via an iterative process. We extracted their R code for the iterative solution as implemented in the \texttt{pmvalsampsize} package, \url{https://github.com/JoieEnsor/pmvalsampsize/blob/main/R/pmvalsampsize_bin.R}, (commit hash a6f867b00dc5a48ce8c3bdc0becc1b40395d06cd) and wrapped it into a custom function using the same number of iterations as the original code (1,000,000). We also wrote R functions implementing the seven closed-form solutions. Note that the code for the iterative method returns the first integer where the confidence interval (CI) width meets or falls below the target CI width for the $C$-statistic. Our R functions rounded up the calculated $N$ to the next integer. 

All seven closed-form functions returned the same values of $N$ shown in Table~\ref{table-1}, given the input $C$, $\phi$ and ${\text{SE}(C)}$ values. Note that the paper itself~\cite{riley_minimum_2021} reported using ${\text{SE}(C)= 0.0255}$, but the supporting information associated with the paper used ${\text{SE}(C)= 0.02551}$, and this one extra decimal place does impact the estimated $N$. For instance, using the parameters from the fifth example of Table~\ref{table-1}, an ${\text{SE}(C)}$ of 0.0255 gives $N = 3274$, whereas an ${\text{SE}(C)}$ of 0.02551 used in their supporting information yields $N = 3271$. We used the latter value of ${\text{SE}(C)}$ for all comparisons to be consistent with the original authors' provided code and examples. Furthermore, Figure~\ref{fig1} illustrates the equivalency of solutions across a wider parameter space, plotting sample size against ${\text{SE}(C)}$ for outcome proportions ($\phi$) ranging from 0.1 to 0.5. The curves across all seven panels are identical, confirming that the distinct closed-form solutions are mathematically equivalent across the domain.
\endgroup

\begin{table}[!ht]
\centering
\caption{Confirmation of closed-form solutions using illustrative examples from Riley et al.~\cite{riley_minimum_2021}. The table shows the parameters ($C$-statistic, $\phi$, ${\text{SE}(C)}$) for five examples reported by  Riley et al.~\cite{riley_minimum_2021}. The resulting sample size ($N$) is identical whether calculated using the original iterative method or any of the seven closed-form solutions. }
\begin{tabular}{lllll}
\toprule
Section$^{a}$ & $C$-statistic & $\phi$ & ${\text{SE}(C)}$ & $N$\\ 
\midrule
3.3.1 & 0.7 & 0.1 & 0.02551 & 1,154 \\ 
3.3.1 & 0.8 & 0.5 & 0.02551 & 302 \\ 
4.3 & 0.8 & 0.018 & 0.02551 & 4,252\\ 
4.3$^{b}$ & 0.75 & 0.018 & 0.02551 & 5,125\\ 
4.3$^{b}$ & 0.85 & 0.018 & 0.02551 & 3,271\\ 
\bottomrule
\end{tabular}
\begin{flushleft} 
\item[$^{a}$] Section of~\cite{riley_minimum_2021} in which the sample size calculation was reported.
\item[$^{b}$] In this section, the $C$-statistic was varied by  Riley et al.~\cite{riley_minimum_2021} as part of a sensitivity analysis.
\end{flushleft}
\label{table-1}
\end{table}

To further verify the robustness of these solutions, we performed a grid-based sensitivity analysis across a wide range of realistic parameters. We calculated the required sample size ($N$) for 450 combinations of $C$-statistic (0.55 to 0.95 in increments of 0.05) and outcome event proportions ($\phi$ = 0.01 to 0.5 in increments of 0.01). In all simulated scenarios, closed-form solutions yielded results consistent with the iterative method (absolute difference $\leq$ 1), without observed numerical instability.

\section*{Benchmarking}
We compared the execution times of the iterative against the closed-form solutions, using the \texttt{microbenchmark} package (version 1.5.0). To capture typical performance and understand the range of execution times due to system variability, each function was evaluated 1,000 times ($\sim{11}$ min to complete). Evaluations were done on a HP Pavilion desktop computer with a AMD Ryzen 7 5700G processor and 16.0 GB RAM running Windows 11 Home (Version 24H2). We acknowledge that the specific results that follow are environment-dependent, and the relative performance differences between functions may vary on other hardware or software configurations. 

\begin{figure}[htbp]
\centering
\includegraphics[width=0.8\textwidth]{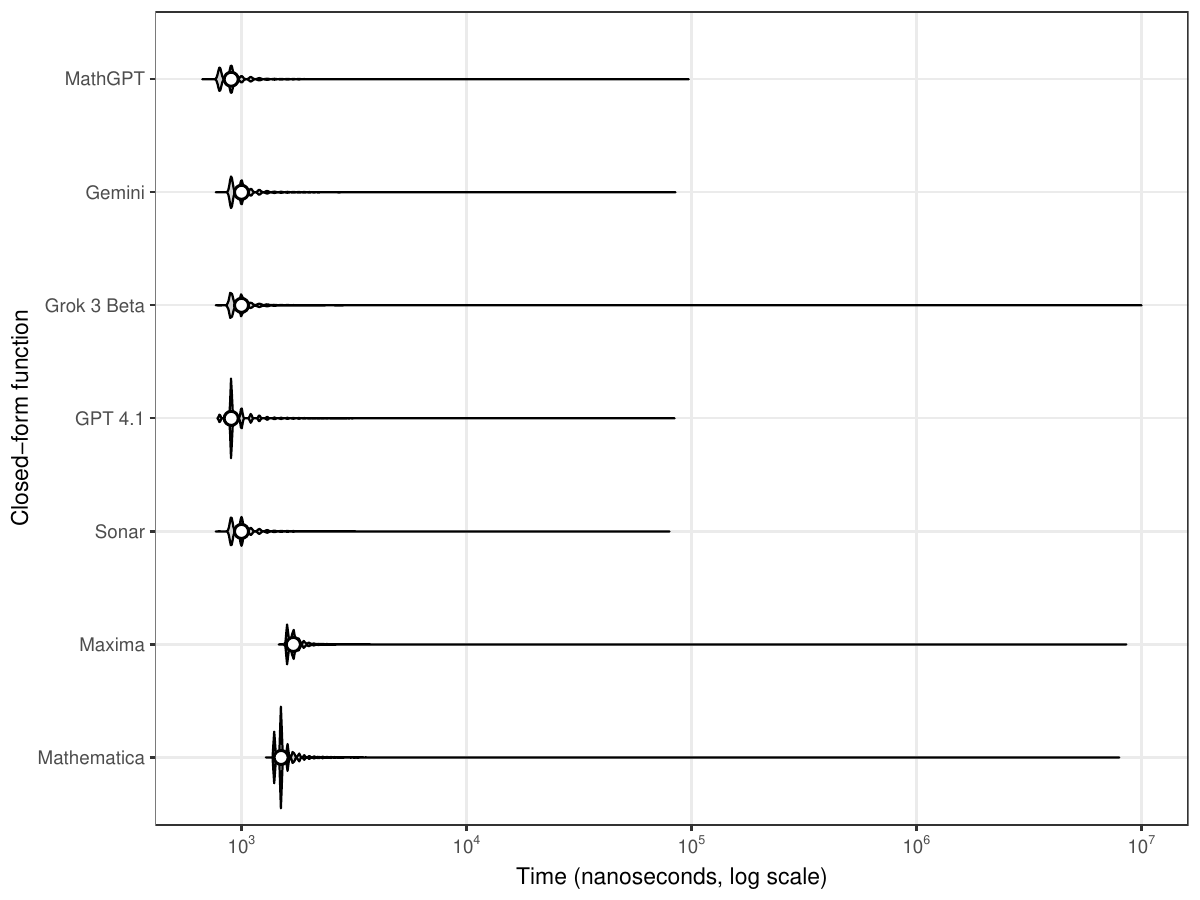} 
\caption{Violin plot of measured execution time (in nanoseconds) of functions implementing closed-form solutions for the sample size required for the external validation of a binary prediction model based on the ${\text{SE}(C)}$ criterion.
Each function was evaluated 100,000 times. Median evaluation times are indicated by the white circles. The width of the shaded area represents the probability density of the data at different values. Note that differences between the seven solutions are negligible on an absolute time scale.}
\label{fig2}
\end{figure}

Based on median execution times, the closed-form solutions were 148,000 to 264,000 times faster than the iterative solution. We then compared the execution times among the seven functions implementing the closed-form solutions, each function being evaluated 100,000 times ($\sim{1.6}$ s to complete). The results are shown in Figure~\ref{fig2}. The AI-derived solutions appeared to be somewhat faster than the CAS solutions in terms of execution time. The code implementations of the GPT 4.1 and MathGPT solutions were the most efficient overall.

\section*{Practical recommendation}
While all seven closed-form solutions are mathematically equivalent, they differ significantly in ease of implementation. We recommend MathGPT's solution for routine use. By defining two intermediate variables ($A$ and $\delta$), this solution reduces the final equation to a compact quadratic form. This modularity minimizes the risk of transcription errors during coding and offers the highest algebraic clarity. Solutions derived via traditional CAS (Mathematica, Maxima), while correct, rely on expanded high-degree polynomials that are opaque to the human eye and more cumbersome to debug.

\section*{Conclusion}
This brief paper presented seven closed-form solutions for directly calculating the minimum sample size required for externally validating a binary prediction model based on controlling the size of the standard error of the $C$-statistic. These solutions yielded identical sample size estimates to a previously published iterative method but were 148,000 to 264,000 times faster. While this speed difference is negligible for a single calculation, the efficiency and stability of closed-form solutions become critical for computationally intensive tasks, such as sensitivity analyses (grid searches) or extensive simulation studies where the function may be called millions of times. 

Although all closed-form solutions were in the same "nanosecond" ballpark of performance, we do reiterate that the derivation process using AI tools is prone to error. AI solutions require careful evaluation. Large language models are designed to generate plausible-sounding text rather than rigorous proofs and can frequently fail at basic arithmetic~\cite{tao_machine_2025,yin_scaffolding_2024}, as we experienced with Claude Sonnet 4.0, GPT 4.1 and Grok 3 Beta. 

We note that the inversion of Newcombe's formula is unique in its complexity among the validation sample size estimation procedures within the Riley et al. framework~\cite{riley_minimum_2021,riley_evaluation_2024}. However, the computational strategy employed here of leveraging CAS and AI to resolve 'intractable' algebra is broadly applicable to any biostatistical problem where mathematical complexity hinders the development of exact solutions.

All R functions implementing the seven closed-form solutions and the iterative comparison are available (\url{https://doi.org/10.5281/zenodo.15850396}). This availability will facilitate direct and more efficient sample size calculations for the external validation of binary prediction models, an aspect rightfully garnering more attention.

\bibliographystyle{unsrt}  
\bibliography{References}  

\end{document}